\documentclass[conference]{IEEEtran}
\usepackage{blindtext}
\usepackage{graphicx}
\usepackage{multirow}
\usepackage[table,xcdraw]{xcolor}
\usepackage{mathtools} 
\usepackage[noadjust]{cite}
\usepackage{array}
\usepackage{float}
\usepackage[letterpaper, margin=0.63in,top= 0.628 in,right=0.64in, left=0.64in]{geometry}

\usepackage{algorithm}
\usepackage[noend]{algpseudocode}
\algrenewcommand\algorithmicrequire{\textbf{Input:}}
\algrenewcommand\algorithmicensure{\textbf{Init:}}
\usepackage[justification=centering]{caption}
\usepackage{graphicx}
\usepackage{subcaption}

\usepackage[table,xcdraw]{xcolor}

\usepackage{cite}

\usepackage{amsfonts}
\usepackage{amsmath}
\usepackage[utf8x]{inputenc}

\usepackage{siunitx}

\usepackage{psfrag}
\usepackage{csquotes}
\usepackage{xspace}

\newcommand{\Xcomment}[1]{}

\makeatletter
\newcommand*{\etc}{%
    \@ifnextchar{.}%
        {etc}%
        {etc.\@\xspace}%
}
\makeatother

\algnewcommand\algorithmicforeach{\textbf{for each}}
\algdef{S}[FOR]{ForEach}[1]{\algorithmicforeach\ #1\ \algorithmicdo}

\begin{document}
\title{Evaluation of Age Control Protocol (ACP) and ACP+ on ESP32}

%or 
%\title{The Real-Time Evaluation of ACP and ACP+ for Static Network}

% Başına ACP32 eklenebilir.
% This report presents

\author{
	{   
	Umut Guloglu\IEEEauthorrefmark{1},
		Sajjad Baghaee\IEEEauthorrefmark{2},
		Elif Uysal\IEEEauthorrefmark{3}
	}	\\
	\IEEEauthorrefmark{1}
	\IEEEauthorrefmark{2}
	\IEEEauthorrefmark{3}
	{Department of Electrical and Electronics Engineering, Middle East Technical University, Ankara, Turkey}\\

	\IEEEauthorrefmark{1}umut.guloglu@metu.edu.tr,
	\IEEEauthorrefmark{2}sajjad@baghaee.com,
	\IEEEauthorrefmark{3}uelif@metu.edu.tr
}

\maketitle
% This file should contain the abstract and keywords

\begin{abstract}

Age Control Protocol (ACP) and its enhanced version, ACP+, are recently proposed transport layer protocols to control Age of Information of data flows. This study presents an experimental evaluation of ACP and ACP+ on the ESP32 microcontroller, a currently popular IoT device. We identify several issues related to the implementation of these protocols on this platform and in general on short-haul, low-delay connections. We propose solutions to overcome these issues in the form of simple modifications to ACP+, and compare the performance of the resulting modified ACP+ with that of the original protocols on a small-delay local wireless IoT connection.

\end{abstract}

%Old one

%Age of Information (AoI) has become an imperative and indispensable metric for time-critical applications. However, protecting the freshness of data is not an easy mission, especially for Internet-of-Things (IoT) applications. The mainstream transport layer protocols, such as TCP and UDP, are not accomplished enough to minimize the AoI. This study presents a real-time evolution of the recently proposed Age Control Protocol (ACP) on ESP32 microcontrollers, one of the various IoT devices. We state the issues related to ESP32 and ACP and bring solutions to solve these successfully. We also compare ACP with another Policy, called Lazy, to see ACP's performance. 


\IEEEpeerreviewmaketitle
\vspace*{-0.5cm}

\section{Introduction}
 In the IEEE special report on Internet of Things issued in March 2014, the term of Internet of Things (IoT) was described as: \enquote{A network of items — each embedded with sensors — which are connected to the Internet.} IoT has applications in numerous domains such as intelligent infrastructures, smart healthcare, smart agriculture, and many more.
 \iffalse
 As IoT covers a wide range of technologies, there are various IoT definitions depending on their applications.
 \fi

%*a network of items where each item is embedded with sensors. Such that these items can collect and exchange data over the Internet with or without human interaction.*%

Remote monitoring, control and automation applications of IoT in areas including, but not limited to, the above domains, have gained importance in recent years.\iffalse These applications tend to have stringent data freshness requirements.\fi The freshness of information updates, such as those involving sensor measurements, is often important to decision-making in computing, accessing, and storing information. For instance, in tracking an object, the latest true and fresh location information of the object enables remote monitoring to estimate the last position accurately \cite{sbaghaee1}.

Age of Information (AoI) \cite{aoi1-1} is a metric characterizing the freshness of an information flow. It is defined as the time that has elapsed since the newest data belonging to this flow, currently available at the destination, was generated at the source. As it is directly related to the freshness of status update based flows, age is gaining popularity as a new Key Performance Indicator (KPI) for machine-type communication systems such as IoT and other status update type applications. The analysis of AoI in queuing models \cite{aoi2-1,aoi2-3} revealed that this metric behaves quite differently from delay. For example, under First-Come-First-Served (FCFS) service, age exhibits non-monotone behaviour with load, i.e., as the load (arrival rate) is increased, age first decreases and then increases. This naturally calls for an optimization, where the packet generation rate can be optimized to keep the load at a point that minimizes age, and more importantly, service policies that favor newer packets. When possible, though, more freedom to this optimization is provided by controlling the packet generation process to respond to the flow's instantaneous age.

Age-optimal generation of update packets for a single-hop network was investigated in \cite{aoi2-4, aoi2-5}, while \cite{aoi2-7} characterized AoI in multiple server systems. A majority of the aforementioned studies focused on time average age. Often, in IoT systems, peak values of age, or age at certain decision instants may be more meaningful as a determinant of performance.
\iffalse
The average peak AoI of multicast transmission with deadlines in IoT networks were investigated in \cite{ Li-multicast-2020}.
\fi
To quantify the freshness decision epochs when the freshness of updates is only important at some decision epochs, age upon decisions (AuD) was introduced in \cite{ Dong-AuD-2019}.
While the theoretical models developed in the literature capture an abstraction of the transport layer to a certain degree, such models and the resulting optimizations require knowledge of network delay statistics, which are not often available in practice.

Recently, a number of efforts have considered AoI in real-life networks \cite{BookAOIP}. An open-source network emulator was used to investigate the AoI in wireless access in \cite{core-aoi}. A similar study in \cite{canberk2018} combined emulation and real-world AoI measurement experiments, reporting AoI measurements for an end-to-end data flows traversing wireless/wired links, affected by various medium access, transport, and network layer scenarios.
Effect of synchronization error on AoI measurement and a solution for calculating an estimate of average AoI without any synchronization requirements has been shown in \cite{8806423}. Authors in \cite{Smart} utilized several transport protocols such as TCP, UDP, and WebSocket to measure AoI on wired and wireless links. Their study includes practical issues such as synchronization and selection of hardware along with transport protocol and their effects on AoI measurements.

Efforts to carry the intuition from theory to practice has included dynamically adaptive approaches such as those that employ Machine Learning methods to age optimization in a network \cite{EgemenRL}. Modifications have been proposed in the transport and application layers to control congestion with an age objective, to keep the \enquote{pipe just full, but not fuller} as suggested in \cite{KleinrockNew}. Recent protocols in that vein include \cite{ bbr, cloudgamingbbr, tanya}.
Shreedhar et al. \cite{tanya} have introduced the Age Control Protocol (ACP) for multi-hop IP networks, enabling timely delivery of updates by adapting the sending rate. They also compared the performance of ACP with the Lazy Policy in their study. The aim of the Lazy policy is to transmit data with an interval around Round Trip Time (RTT). An improved version of ACP, named ACP+, was presented in \cite{shreedhar2021empirical}. According to \cite{shreedhar2021empirical}, ACP+ has superior performance for timely delivery of updates over fat pipes and long paths.

The goal of this paper is to report experimental test results of ACP and ACP+ over shorter paths, and low rates, in an IoT scenario, and to suggest how the protocol may be modified for this scenario. The outline of the rest of the paper is as follows. Section \ref{aoi-intro} provides the definition of age and an expression for the computation of its time average on a given packet trace. Descriptions of ACP and ACP+ and their distinctions are presented in Section \ref{acp_tanitim}. Section \ref{emu-testbed} describes the testbed used in our study. Section \ref{problem_solution} provides a discussion related to the issues and solutions of using ACP and ACP+ on the ESP32 microcontroller. Results and conclusions are presented in Section \ref{results} and \ref{conclusion}, respectively.

%*The emulation testbed is described in  Section  \ref{emu-testbed}, and the real-world testbed is introduced in section  \ref{physical-testbed}.  Results obtained on these testbeds for TCP/IP running on various physical links such as WiFi, Ethernet, LTE, 3G  and  2G  are presented  in  Section  \ref{results}.  Some conclusions and future directions are discussed in section \ref{conclusion}.*%

\vspace{-0.085in}
\section{AoI: definition and computation}
\label{aoi-intro}

\begin{figure}
	\centering
	\includegraphics[width=0.84\linewidth]{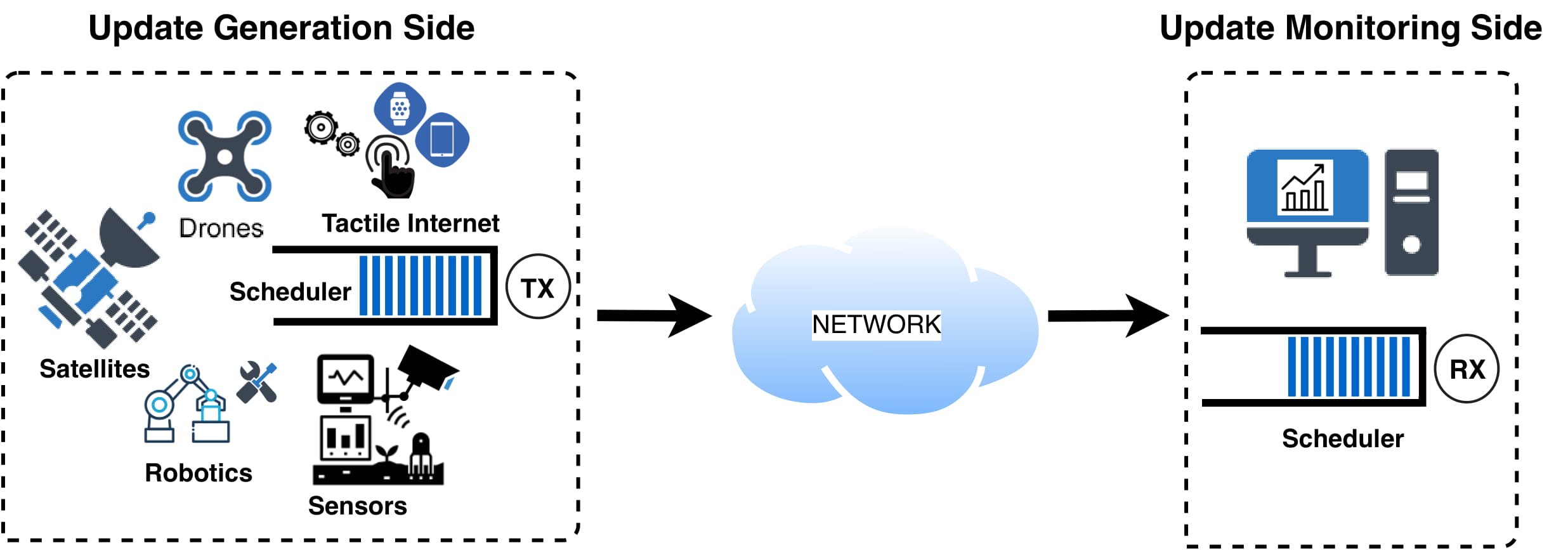}
		\vspace{-0.05in}
	\caption{A status update communication setup.}
	\label{aoiCommunicationSetup}
\vspace{-0.1in}
\end{figure}

Consider a status update scenario (Fig.\ref{aoiCommunicationSetup}), where a process running at a destination node needs samples (status updates) from a remote source. After the generation of an update at timestamp $U(t)$ at the source, that update will start aging. When the receiver is not fed with fresh updates, the process at the receiver may suffer due to stale data. Once a fresh update is received, the age on the monitoring side is updated with the age of the recently arrived update. The status age, $\Delta(t)$, is calculated as $\Delta(t) = t - U(t)$, where $t$ is the time of the newest update available at the monitoring side.

\begin{figure}
	\centering
	\includegraphics[width=0.84\linewidth]{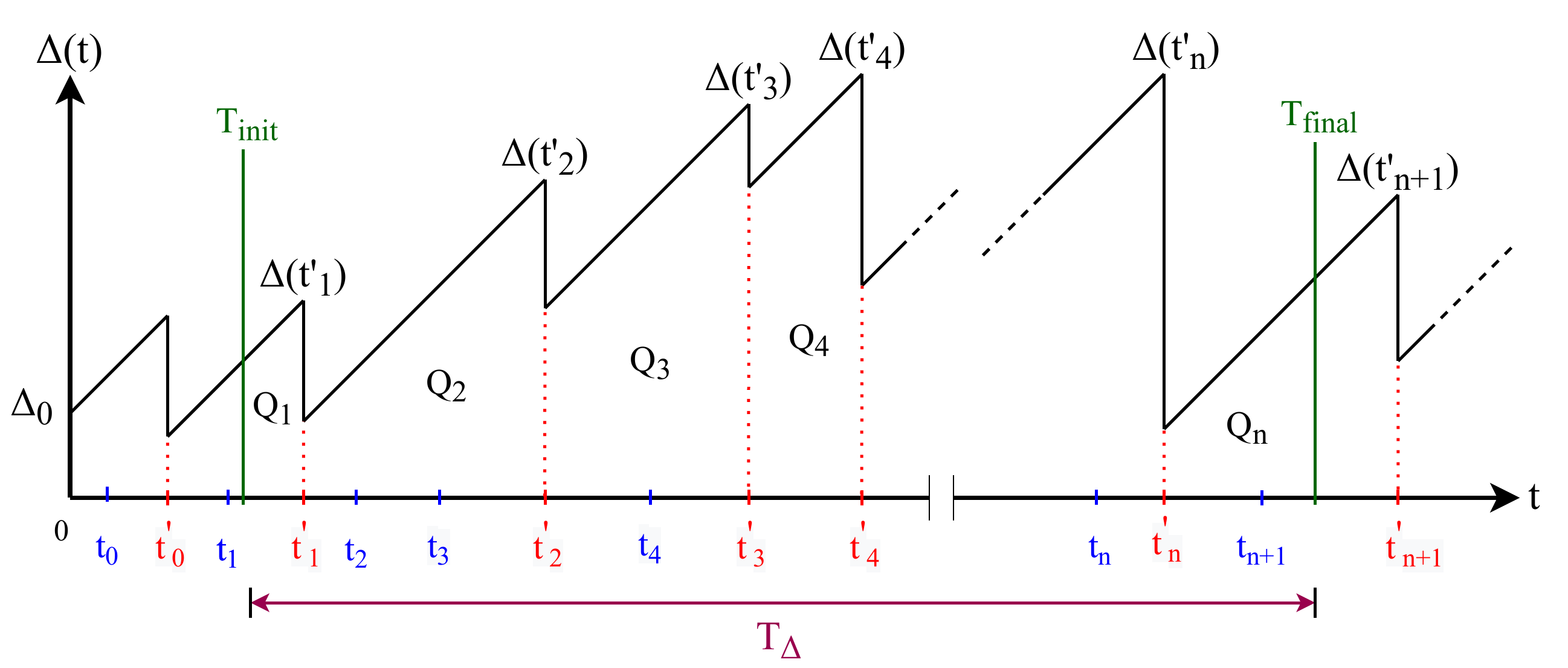}
	\vspace{-0.12in}
	\caption{Sample path of the age process: $\Delta(t)$}
	\label{aoi1}
	\vspace{-0.25in}
\end{figure}

Based on this definition, age is a continuous-time continuous-valued process with sample paths that follow a sawtooth pattern (Fig. \ref{aoi1}). Without loss of generality, let us assumed that the observation begins at $t=0$ and update generation starts at $t>0$ with empty transmitter and receiver queues. The initial age is $\Delta(0) = \Delta_{0}\geq 0$. The source generates status updates at $t_{0}$, $t_{1}$, $\cdots$, $t_{n}$, which reach the monitor at $t'_{0}$, $t'_{1}$, $\cdots$, $t'_{n}$.

\iffalse
\begin{equation}
	%\vspace{-0.12in}
\Delta(t) = t - U(t)
\label{eq:Statusaoi}
\end{equation}
\fi

The area under the age graph within a time window of length $T_\Delta=T_{final}-T_{init}$, (Fig. \ref{aoi1}) normalized by $T_\Delta$ provides the time average age within this window:
\vspace{-0.013in}
\begin{equation}
\vspace{-0.013in}
\overline{\Delta} = \frac{1}{T_\Delta}\int_{T_{init}}^{T_{final}} \Delta(t) dt =  \frac{\sum_{i=1}^{n} Q_i}{T_\Delta}
\label{eq:aoi}
\end{equation}

\iffalse
In practice,
consider a packet trace including $n$ updates. The area under age graph, (Fig. \ref{aoi1}), is made up of the trapezoidal areas $\{Q_i\}$ for $1\leq i \leq n$. By substituting $Q_i$'s in (\ref{eq:aoi}), $\overline{\Delta}$ will be as (\ref{eq:age with q}), where ~\eqref{eq:age with q1},\eqref{eq:age with q2} and \eqref{eq:age with q3} provide $Q_1$, $Q_i$ for $2\leq i \leq n-1$ and $Q_n$, in terms of the update generation and reception times.
\fi

In practice, consider a packet trace including $n$ updates and age graph as shown in Fig. \ref{aoi1}. The area under this graph is made up of $n$ trapezoidal areas, where $Q_1$, $Q_i$ for $2\leq i \leq n-1$ and $Q_n$, are calculated as ~\eqref{eq:age with q1},\eqref{eq:age with q2} and \eqref{eq:age with q3}.

\begin{table}[ht]
\vspace{-0.2in}
\begin{center}
\caption {Decision algorithm of ACP}
\normalsize
\vspace{-0.1cm}

\begin{tabular}{ |c|c||c| }
    \hline
    $\delta_k$ & $b_k$ & ACP targets to: \\
    %$\overline{\Delta}_k-\overline{\Delta}_{k-1}$ & $\overline{B}_k-\overline{B}_{k-1}$ & ACP targets to: \\
    \hline
    $>0$ & $>0$ & Decrease the backlog \\  
    $>0$ & $<0$ & Increase the backlog \\
    $<0$ & $>0$ & Increase the backlog \\
    $<0$ & $<0$ & Decrease the backlog \\
    \hline
\end{tabular}
\label{tab:algo}
\end{center}

\end{table}

%\begin{equation}
%\vspace{-0.12in}
%\overline{\Delta} = %\frac{\sum_{i=1}^{n} Q_i}{T_\Delta}
%\label{eq:age with q}
%\end{equation}
\vspace{-0.1in}  %1.25
\begin{equation}
\vspace{-0.125in}
Q_1=\frac{1}{2} (T_{init}+t'_{1}-2t_0)(t'_{1}-T_{init})
\label{eq:age with q1}
\end{equation}
\vspace{-0.125in}
\begin{equation}
\vspace{-0.125in}
Q_i = \frac{1}{2} (t'_{i}+t'_{i-1}-2t_{i-1})(t'_{i}-t'_{i-1})
\label{eq:age with q2}
\end{equation}

\begin{equation}
\vspace{-0.125in}
Q_n=\frac{1}{2} (T_{final}+t'_{n}-2t_n)(T_{final}-t'_{n})
\label{eq:age with q3}
\vspace{0.1in}
\end{equation}

\section{Age Control Protocol (ACP) and ACP+}
\label{acp_tanitim}

%Both ACP and ACP+ are transport layer protocols designed to reduce the age \cite{tanya},\cite{shreedhar2021empirical}. They are nearly the same but with minute distinctions. In this section, we explain how ACP works and explicate the distinctions between ACP and ACP+. 

% This is the new one
ACP and ACP+ are transport layer protocols designed to minimize the age of information at the monitor (receiver) by altering the updating (sending) rate of the source (sender) according to the estimated network conditions. In this section, we describe ACP and the differences between ACP and ACP+. 

ACP is built on the User Datagram Protocol (UDP), and operates on each end host of a session and consists of two phases: the initialization phase, and the \enquote{epochs} phase. In the initialization phase, the source sends a certain number of packets to the monitor, waits for the ACK packets, and computes a Round-Trip Time (RTT) estimate based on these. The purpose of this phase is to configure the initial update rate of the source. 
%Time synchronization problem between sender and receiver can be solved in this phase. 
%After the initialization phase, the protocol enters the second phase, which is the epochs phase. 
This phase is followed by the epochs phase, divided into
%a total of $N$ numbered 
periods called epochs. In each epoch, labeled with $k$, the average AoI ($\overline{\Delta}_k$) and average backlog ($\overline{B}_k$) (average number of packets sent to the monitor, but not yet acknowledged) are computed. ACP decides to increase or decrease the backlog according to changes in $\overline{\Delta}_k$ and $\overline{B}_k$. The decision algorithm of the protocol is shown in Table \ref{tab:algo}, where $\delta_k$ is defined as $\overline{\Delta}_k-\overline{\Delta}_{k-1}$ and $b_k$ is defined as $\overline{B}_k-\overline{B}_{k-1}$. %The algorithm simply tries to reduce AoI by using trial-and-error method. 

When an ACK packet arrives at the source, the exponentially weighted moving average (EWMA) of RTT ($\overline{RTT}$), and the EWMA of inter-ACK arrival times ($\overline{Z}$) are calculated. ACP uses these to infer network conditions and set an appropriate update rate ($\lambda_k$), which will be used during the next epoch. By changing $\lambda_k$, the protocol can alter the backlog and thereby control the age. For that purpose, ACP chooses an action and aims to change the average backlog by $b_{k+1}^*$ packets according to changes in $\overline{B}_k$ and $\overline{\Delta}_k$, where $b_{k+1}^*$ is the desired backlog change. The computation of the update rate is given by (\ref{eq:update_rate}). 

At each epoch, one of the following three actions, described in Table \ref{tab:actions}, are chosen: \enquote{INC}, \enquote{DEC}, and \enquote{MDEC}. Each action results in a certain corresponding $b_{k+1}^*$. The value of step-size parameter $\kappa$, see Table \ref{tab:actions}, is constant during packet transmission. It is crucial for ACP as the actions \enquote{INC} and \enquote{DEC} aim to change the average backlog by $\kappa$ packets.
%The value of $\kappa$ is constant during packet transmission.
ACP uses the \enquote{MDEC} action to obtain a multiplicative decrease when \enquote{DEC} does not  decrease $\overline{B}_k$ sufficiently fast. 

\vspace{-0.1in}
\begin{equation}
\lambda_k = \frac{1}{\overline{Z}} + \frac{b_{k+1}^*}{min(\overline{RTT},\overline{Z})}
\label{eq:update_rate}
\vspace{-0.075in}
\end{equation}

ACP also assigns the length of each epoch ($\overline{T}$) using (\ref{eq:epoch_length}) at epoch boundaries. The epoch duration must be short enough to respond to the changes promptly and long enough to provide an accurate estimate of network conditions. 
%\vspace{-0.03in}
%By doing that, ACP tries to protect data's freshness by adapting to the network changes \cite{tanya}. The aggressiveness of the protocol is controlled by a predefined constant parameter $\kappa$. 

\begin{table}
\vspace{-0.1in}
\begin{center}
\caption {Table of actions and corresponding $b_{k+1}^*$ values }
\normalsize
\vspace{-0.25cm}

\begin{tabular}{ |c||c| }

    \hline
    The Action & $b_{k+1}^*$ \\
    \hline
    Increase (INC) & $\kappa$   \\
    Decrease (DEC) & $-\kappa$  \\
    Multiplicative Decrease (MDEC) & $-(1-2^{-\gamma})B_k$ \\
    \hline
    
\end{tabular}
%\vspace{0.2in}
\label{tab:actions}

%INC" and "DEC" are the general actions aiming to increase/decrease the average backlog $b_{k+1}^*$ packets. "MDEC" is used when the power "DEC" is not enough to achieve lower AoI fast.  

%
\end{center}
\vspace{-0.1in}
\end{table}

\vspace{-0.09in}
\begin{equation}
\vspace{-0.05in}
\overline{T} = 10 \times min(\overline{RTT},\overline{Z})
\label{eq:epoch_length}
\end{equation}

ACP+, an enhanced Age Control Protocol, is based on ACP, with several modifications:

\begin{itemize}
  \item \textbf{Computation of $\lambda_k$:} While ACP uses (\ref{eq:update_rate}) to update $\lambda_k$, ACP+ uses a slightly altered equation where $min(\overline{RTT},\overline{Z})$ is replaced by $\overline{RTT}$.

  \item \textbf{Setting $\kappa=1$ and clamping $\lambda_k$:} The other substantial difference is that $\kappa$ is set to $1$ in ACP+, and $\lambda_k$ is clamped in order not to alter the update rate dramatically. In \cite{shreedhar2021empirical} authors specified the clamping boundaries as $1.25\times\lambda_{k-1}$ (Max update rate) and $0.75\times\lambda_{k-1}$ (Min update rate) in which $\lambda_{k-1}$ stands for the update rate of the previous epoch.
 %If $\lambda_k>1.25\times\lambda_{k-1}$ , then ACP sets $\lambda_k$ to $1.25\times\lambda_{k-1}$, and if $\lambda_k<0.75\times\lambda_{k-1}$, the update rate $\lambda_k$ is set to $0.75\times\lambda_{k-1}$.
  
  \item \textbf{Modification in Eq. (\ref{eq:epoch_length}):} 
  %The last distinction is in the equation for designating $\overline{T}$. 
  ACP+ uses (\ref{eq:epoch_length+}) to set $\overline{T}$. This is obtained by replacing $min(\overline{RTT},\overline{Z})$ in (\ref{eq:epoch_length}) by $\frac{1}{\lambda_k}$. This roughly corresponds to ACP+ targeting to send about $10$ packets over an epoch duration. 

\vspace{-0.05in}
\begin{equation}
\overline{T} = \frac{10}{\lambda_k}
\label{eq:epoch_length+}
\end{equation}
  
\end{itemize}

%We observed that altering clamping boundaries
%and the equation of epochs' length 
%makes noteworthy improvements in terms of age in a small-delay network. The detailed report can be found in Section \ref{results}.  

%In this study, we have used ESP32's to send and receive data in a 1-hop network. While implementing ACP, we have encountered several problems which might change the success of the protocol significantly. Next, we will see these problems and their possible solutions. 
\vspace{-0.0in}
\section{The Testbed}
\label{emu-testbed}

\begin{figure}[ht]
	\centering
	\includegraphics[width=0.84\linewidth]{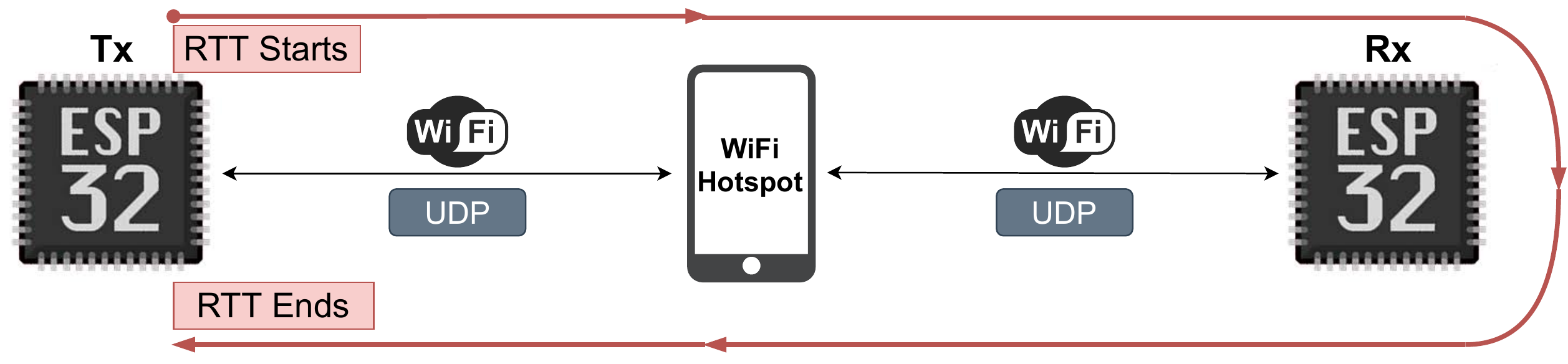}
		\vspace{-0.1in}
	\caption{The schematic of the testbed}
	\label{testbedsema}
\vspace{-0.2in}
\end{figure}

Our motivation is to determine whether ACP and ACP+, which were successfully tested on long-haul, high rate connections, also successfully minimize the age of information in small-delay real-world IoT networks, and if not, explore whether certain modifications can be proposed to better cater for these settings.  For this purpose, we designed a test setup (Fig. \ref{testbedsema}), which contains two ESP32 low-power systems with on chip microcontrollers and integrated Wi-Fi, and a cellular phone used as a Hotspot. In our setup, one of the ESP32 devices is assigned as a source (TX), whereas the other is appointed as a monitor (RX). The transmitter node sends packets using UDP. By listening to the serial port of TX, data logging in each epoch is done. It follows by visualizing and analyzing the results via MATLAB software. %New

%The Arduino serial port was used to obtain the calculated average AoI in each epoch, and MATLAB was used to visualize the data.

In this experiment, ACP/ACP+ are implemented in the transmitter node. The transmitter sends packets, and the receiver reflects the incoming packet back to the sender in place of an ACK. The transmitter node calculates all necessary information related to ACP/ACP+ and decides the action which must be taken. The receiver node acts as an echo server in this experiment.

We primarily compared ACP with ACP+ and the Lazy Policy in \cite{tanya}. The Lazy Policy essentially tries to keep the sending period around RTT, hence the update rate is set to $\frac{1}{\overline{RTT}}$ at the end of each epoch. This policy tries to keep the backlog around $1$ following the guidelines given in \cite{KleinrockNew}, and was used as a simple benchmark following \cite{tanya}. 

%We primarily compared ACP with ACP+ and the Lazy Policy in \cite{tanya}. The Lazy Policy essentially tries to keep the sending period around RTT, hence the update rate is set to $\frac{1}{\overline{RTT}}$ at the end of each epoch. This policy tries to keep the backlog around $1$ following the guidelines given in \cite{KleinrockNew}, and was used as a simple benchmark following \cite{tanya}. 

%We primarily compared ACP with ACP+, and Lazy Policy \cite{tanya} for different $\kappa$ values. Lazy Policy tries to keep the sending period around RTT, as it has determined the backlog to be kept around $1$ \cite{tanya}. 

% NEW
%For this reason, the update rate is set to $\frac{1}{\overline{RTT}}$ at the end of each epoch. All other calculations related to ACP or ACP+ are done again. For instance, Lazy still calculates $\overline{B}_k$, $\overline{Z}$, chooses the action, and designates $\lambda_k$. Thus, we make the computational costs of all protocols almost equal. 
%For this reason, the update rate is set to $\frac{1}{\overline{RTT}}$ at the end of each epoch. 
%In the experiment, we run 5 times for Lazy Policy, ACP+, and ACP for 8 different $\kappa$ values.
A set of 5 different experiments were carried out, running Lazy, ACP+, and ACP side by side. In each experiment, there were 5 runs for each of the 8 different $\kappa$ values. In each run, the transmitter sent 10.000 packets to the receiver. Each packet had a 4-byte payload. 
%In each run, the transmitter node sent 10.000 packets, each of which has a 4-bytes payload, to the receiver node and estimated average AoI. 
%Indeed, we modified ACP+ by altering clamping boundaries. Normally original ACP+ clamps $\lambda_k$ if it is bigger than $1.25 * \lambda_{k-1}$ or smaller than $0.75*\lambda_{k-1}$. In the experiments, we noticed that ACP+ clamps approximately 81\% of the designated $\lambda_k$ values and realized the boundaries of clamping are far from $\lambda_{k-1}$, which makes ACP+ a greedy protocol for static networks. 

In the second set of experiments, we focused on improving ACP+ for small-delay networks and testing this modified (improved) version. The original ACP+ clamps $\lambda_k$ if it is bigger than $1.25 \times \lambda_{k-1}$ or smaller than $0.75\times \lambda_{k-1}$. During the experiments, we noticed that ACP+ clamps approximately 81\% of the designated $\lambda_k$ values and realized the boundaries of clamping are far from $\lambda_{k-1}$, which makes ACP+ a greedy protocol for small-delay networks. Therefore, we modified ACP+ by altering clamping boundaries as $1.1 \times \lambda_{k-1}$ and $0.9 \times \lambda_{k-1}$ and ran the test 5 times for the original and modified ACP+ versions with two different epoch length scenarios. These scenarios are detailed in Section \ref{problem_solution}. We also ran ACP with $\kappa$ value giving the best result in terms of AoI to compare with the new ACP+ version. The estimated average $\overline{\Delta}_k$ values under the original and modified ACP+ versions are reported in Table \ref{tab:modify_tablo}. Lastly, we traced the age over time for ACP+ and Lazy Policy and tested whether the feedback mechanism, see Sec. \ref{problem_solution}, is successful in peak age violation cases.

%First, the sender ESP32 sent data for approximately 90 seconds, and the graph of AoI for both ACP and Lazy is observed as seen in Fig. \ref{agetime}. Then, we tested ACP for different $\kappa$ values as it is possibly the most important parameter of the protocol. It controls the aggressiveness of the protocol. \cite{tanya} states that $\kappa = 0.25$ gives the best results for simulations, and $\kappa= \{1 , 2\}$ is more suitable for intercontinental data transfer. In our test setup, we tested 12 different $\kappa$ values and Lazy Policy. We call a set when the sender ESP32 sends 10,000 messages to the receiver and receives the ACK messages of them. For each $\kappa$ value and Lazy Policy, we did our experiment by using 5 sets.

%In some epochs, the peak age was violated due to Problem 1 and Problem 3. Fig. \ref{yesNeglect} shows the graph of average AoI versus $\kappa$ values when the age values exceeding the peak age threshold value are neglected (they are not used in the calculations), whereas Fig. \ref{noNeglect} shows the same graph when the age values exceeding the peak age threshold value are not neglected (they are used in the calculations). In addition, Fig. \ref{Median} shows the graph of median AoI versus $\kappa$ values. As the peak age violations are relatively rare, the median AoI did not change with/without neglecting the violation values.

\vspace{-0.00in}
\section{Discussions}
\label{problem_solution}

% Hafif bir özet yap.

Several issues that we have recognized to possibly affect the success of the protocols under study will be discussed next. This will be followed by suggested solutions. Some of the issues are due to the particular device hardware (which we believe will be typical of other IoT devices), while others are due to the parameter settings of ACP, or ACP+. 
	\vspace{-0.1in}
\subsection{Issues}

\begin{itemize}
\item  \textbf{Issue 1: Unnecessary Queuing Delay in ESP32’s}
\end{itemize}

While sending and retrieving the packets, it is noticed that the measured RTT values are larger than the expected ones on occasion. This issue has been encountered in independent work \cite{Smart}, which suggested an underlying buffer management error for its cause. It is observed that ESP32 starts to process the incoming packet after several packets arrive at the buffer.

\begin{itemize}
\item  \textbf{Issue 2: Error in Average Age Calculation}
\end{itemize}

%The average age of each epoch is calculated by dividing the area under the age graph by the epoch's length. The problem is the time interval between the arriving time of the last received data and the ending, or starting, time of the epochs. These areas, by definition, must be added to the calculations resulting in the changes in average age. These changes might affect the action of ACP and ACP+ according to the algorithm of the protocols.

The average age in each epoch is calculated by dividing the area under the age graph by the duration of the epoch. However, there is a fringe effect due to the time interval between the arriving time of the last received data and the ending, or starting, time of the epochs, i.e., $Q_1$ and $Q_n$ in Fig. \ref{aoi1}. 
%These areas, by definition, must be added to the calculations resulting in the changes in average age. 
This results in a small overestimation of average age, which may incorrectly alter the chosen action of ACP and ACP+ in the case of small epoch durations.
%\newpage
\begin{itemize}
\item  \textbf{Issue 3: Failure of The System Due to Issue 1}
\end{itemize}

%If the system is overly affected because of Problem 1, the system might enter a loop hard to exit. When a message waits a long time in the queue, there will be an increase in RTT and inter-update arrival time values resulting in a decrease in update rate. Suppose the same problem, Problem 1, continues. In that case, the system cannot increase the update rate anymore since not increasing the update rate is the right action according to the measured RTT and inter-update arrival time values. However, there is a plausibility that these values are wrong due to Problem 1, as we supposed.

Assume Issue 1 occurs, and packets incur a long delay in the queue. In this case, $\overline{RTT}$ and $\overline{Z}$ increase, which triggers the system to decrease $\lambda_k$. Normally, the \enquote{INC} action increases $\overline{B}_k$ by $\kappa$ packets according to (\ref{eq:update_rate}), if the calculated $\overline{RTT}$ and $\overline{Z}$ are true values. However, in our case, ACP estimates the network conditions like an intercontinental network and sends data with a slow rate. If the problem continues, it always overestimates $\overline{RTT}$ and $\overline{Z}$, and cannot increase the update rate to the optimal levels since it is the right move according to the algorithm. Thus, it enters a loop that is tough to exit. 

%\begin{itemize}
%\item  Problem 4: Negative Update Rate
%\end{itemize}

%As mentioned earlier, the update rate is calculated by formula \ref{updaterate}. It can be noticed that this update rate can take negative values when the action is "DEC" or "MDEC", i.e., if $b_{k+1}^*$ is negative, $\lambda_k$ might be negative as well.

\vspace{-0.1in}
\subsection{Proposed Solutions}

\begin{itemize}
\item  \textbf{Solution 1: Increasing the Length of the Epochs}
\end{itemize}

Increasing the length of the epochs is a solution for Issue 1 and Issue 2. In \cite{tanya}, \cite{shreedhar2021empirical}, the epoch length is calculated by  (\ref{eq:epoch_length}) and  (\ref{eq:epoch_length+}) for ACP and ACP+, respectively. This can be called  $\mathbf{\overline{T}_{10}}$ case. In our design, we defined the length as $30 \times min(\overline{RTT},\overline{Z})$ for ACP and $\frac{30}{\lambda_k}$ for ACP+ by aiming to increase the length. This case is named $\mathbf{\overline{T}_{30}}$. Except in the test to compare these two cases, the $\mathbf{\overline{T}_{30}}$ case is used in the rest of this study.

Suppose Issue 1 is occasionally encountered. In that case, an increase in the age and thereby an abnormal increase in the area under the AoI curve is expected. By increasing the epoch length sufficiently, the effect of this temporary increase may be mitigated. Of course, as the length is increased, the system reacts more slowly than the one with a shorter length case to sudden packet traffic variations and RTT changes. We tested the outcomes of increasing the epoch length. According to the results (Section \ref{results}), it is deduced that even if this is a considerable issue in dynamic networks, it is beneficial for small-delay networks like the one under consideration.

\begin{itemize}
\item  \textbf{Solution 2: Adding a Feedback Mechanism}
\end{itemize}

This is a solution for Issue 3, and it can only be used in one-hop networks with one client and one server. It is designed for test purposes. The update rate must be increased to exit from the loop caused by Issue 3. If ACP increases the update rate, the buffer is filled in a shorter time. Thus, the difference between the measured age value, which is not correct due to Issue 1, and the actual age value decreases. For this purpose, we designed a feedback algorithm that will be active when the age exceeds a predefined peak age threshold value. In our experiments, $200$ ms is set as the peak age threshold. In this algorithm, the minimum value of RTT of each epoch is calculated. If the peak age is violated, the feedback mechanism is activated, and the new $\overline{RTT}$ is calculated by  (\ref{eq:feedback}).

\begin{equation}
\vspace{-0.08cm}
\overline{RTT} = \frac{\overline{RTT}+\zeta\times RTT_{min,epoch}}{\zeta +1}
\label{eq:feedback}
\vspace{+0.05cm}
\end{equation}

$\zeta$ in (\ref{eq:feedback}) is set to $0$ initially. After each epoch, the mechanism checks whether there exists a peak age violation or not. If the peak age is violated, $\zeta$ is increased by $1$, and it becomes $0$ otherwise. In other words, if the violation occurs consecutively, the aggressiveness of the mechanism increases. As a result, $\overline{RTT}$ decreases considerably, which leads to an increase in the update rate. The pseudocode of the mechanism is shown in Algorithm
\ref{Algo}.
	\vspace{-0.05in}
	
\begin{algorithm}
\caption{Feedback Mechanism for Test Measurements}
\begin{algorithmic} 
\Require $\overline{\Delta}_k ,\overline{T}$ and $\zeta \leftarrow 0$
%\Ensure $\zeta \leftarrow 0$
\While{$ true $}
\If{$\overline{\Delta}_k > Peak \ Age \ Threshold$}
\State $\zeta = \zeta +1$
\Else 
\State $\zeta \leftarrow 0$
\EndIf
\State $update \ \overline{RTT}$ and wait  $\overline{T} $
\EndWhile
\end{algorithmic}
\label{Algo}
\end{algorithm}
%	\vspace{-0.1in}
%The aim of this feedback mechanism is to exit from the loop and return to the initial position, which is the position before Problem 1 happens, assuming there will be no problem related to ESP32 queues.
This feedback mechanism aims to exit from the loop and return to the optimum update rate's vicinity. As explained previously, it is designed for peer-to-peer communications over a one-hop system. If there is more than one source or more than one monitor, or another source sending data using the same access point, it is not possible to know whether Issue 1 or the traffic congestion is the reason for the violation. 
%
%\begin{itemize}
%\item  Solution 3: Adding a Fallback Mechanism Changing %Update Rate to Code
%\end{itemize}

%This is the solution for Problem 3. 

%\vspace{-0.12in}
\section{Results}
\label{results}

%In some epochs, the peak age was violated due to Problem 1 and Problem 3. Fig. \ref{yesNeglect} shows the graph of average AoI versus $\kappa$ values when the age values exceeding the peak age threshold value are neglected (they are not used in the calculations), whereas Fig. \ref{noNeglect} shows the same graph when the age values exceeding the peak age threshold value are not neglected (they are used in the calculations). In addition, Fig. \ref{Median} shows the graph of median AoI versus $\kappa$ values. As the peak age violations are relatively rare, the median AoI did not change with/without neglecting the violation values.

%Experiment results have been demonstrated and discussed in this section.

\begin{figure} [ht]
\centering
\begin{subfigure}{0.48\textwidth}
    \centering
	\includegraphics[width=0.84\linewidth]{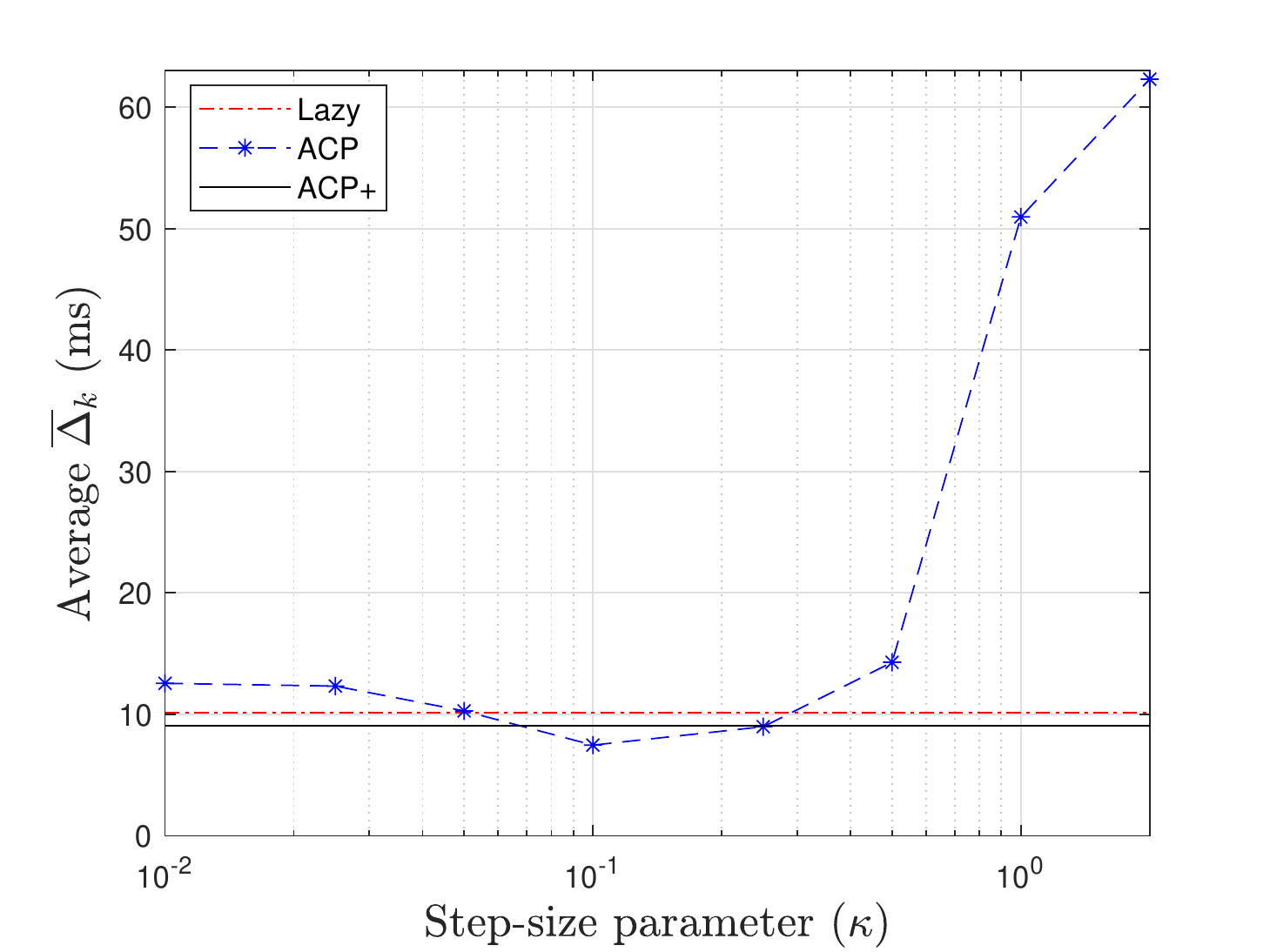}
%	\vspace{-0.1in}
	\caption{The empirical average $\overline{\Delta}_k$ results}
	\label{average}
\end{subfigure}
	\vspace{-0.2in}

\begin{subfigure}{0.48\textwidth}
    \centering
    \includegraphics[width=0.84\linewidth]{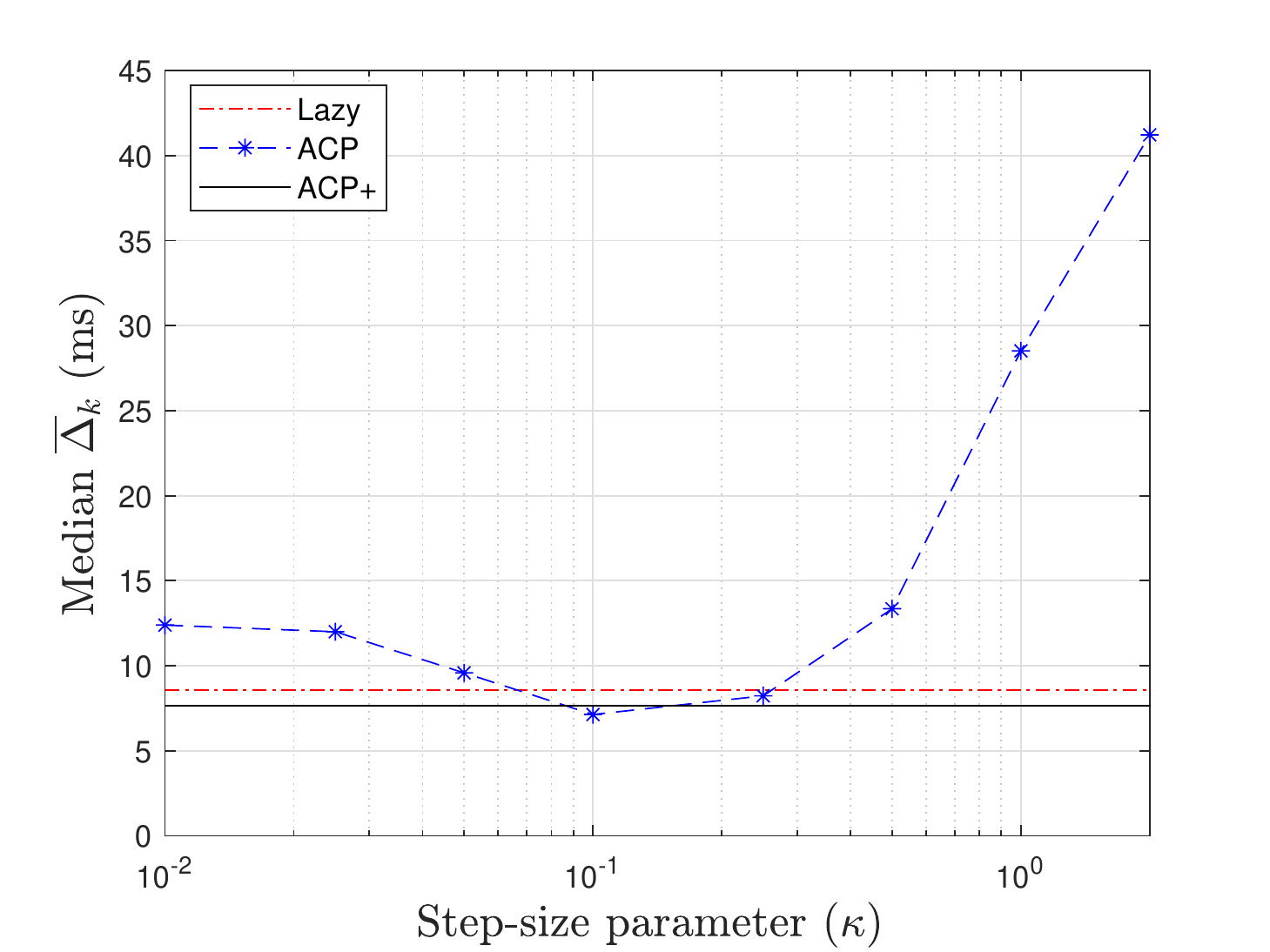}
%    	\vspace{-0.1in}
	\caption{The empirical median $\overline{\Delta}_k$ results}
	\label{median}
\end{subfigure}
\vspace{-0.2in}
\caption{The empirical $\overline{\Delta}_k$ results for Lazy Policy, ACP with different $\kappa$ values and ACP+}
\vspace{-0.2in}
\end{figure}

\begin{figure}
	\centering
	\includegraphics[width=0.84\linewidth]{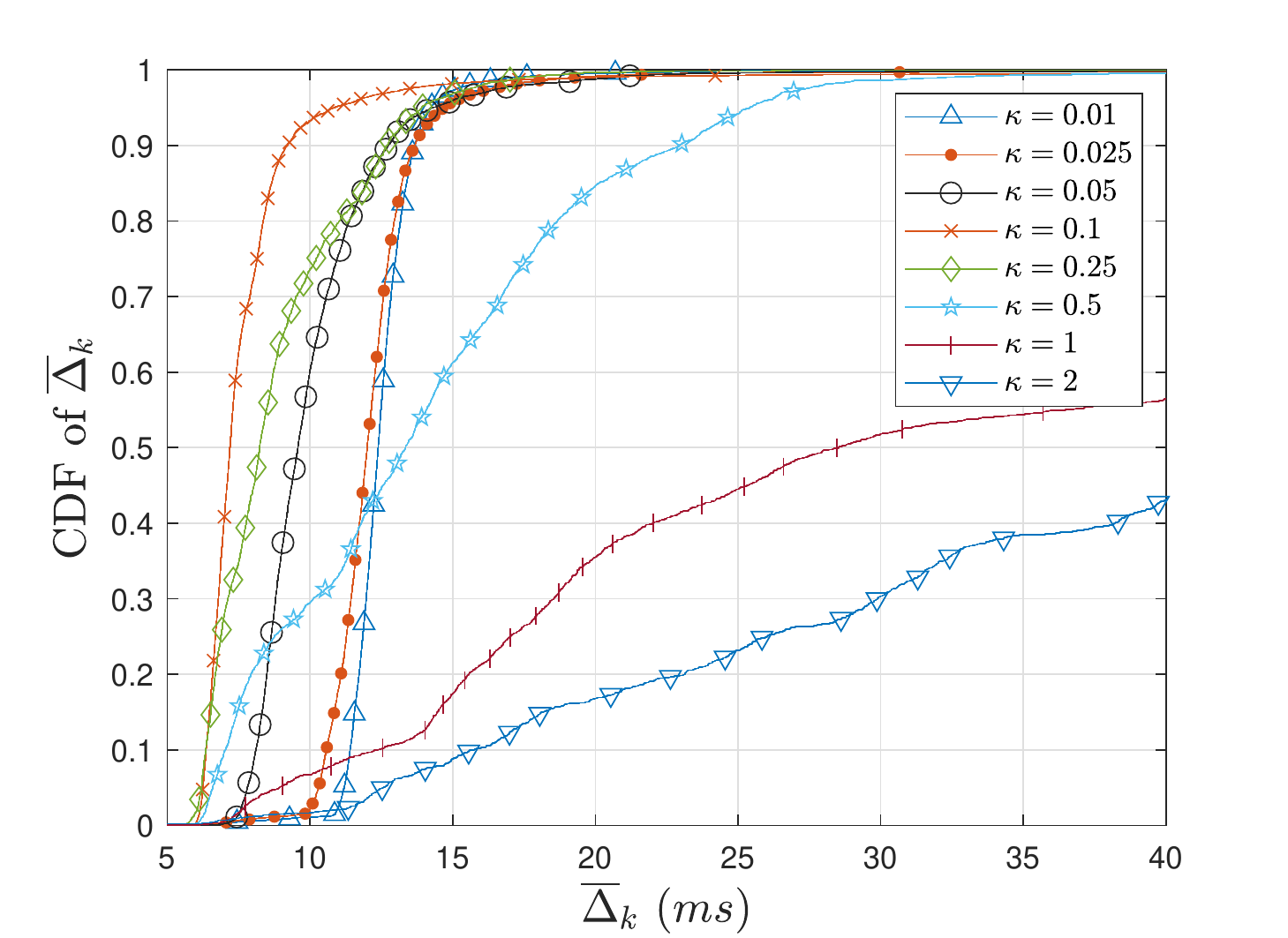}
	\vspace{-0.1in}
	\caption{The empirical AoI CDFs for ACP with different $\kappa$}
	\label{kappa_cdf}
	\vspace{-0.2in}
\end{figure}

To examine the effect of the step-size parameter ($\kappa$) on AoI in ACP, a series of experiments have been conducted. Fig. \ref{average} and Fig. \ref{median} compare ACP with different $\kappa$ values, original ACP+ and Lazy Policy in terms of empirical $\overline{\Delta}_k$ (average age over an epoch). Experimentally, ACP with $\kappa=0.1$ is the best from the point of average $\overline{\Delta}_k$, see Fig. \ref{average}, and median $\overline{\Delta}_k$, see Fig. \ref{median}. Smaller $\kappa$ values lead to higher age as ACP cannot react in time. On the other hand, larger values of $\kappa$ also lead to higher age because ACP becomes over-greedy, causing redundant $\lambda_k$ oscillations. The cumulative distribution function (CDF) of $\overline{\Delta}_k$ under ACP protocol is depicted in Fig. \ref{kappa_cdf}. The reason for the flatter curves in the $\kappa=\{0.5,1,2\}$ cases is the unnecessary oscillations as stated before. As a result, it appears that $\kappa$ must be chosen with care to obtain optimal results.  However, as noted in \cite{shreedhar2021empirical}, $\kappa$ is a parameter difficult to set in general. ACP with the same $\kappa$ value might work differently in different time intervals and networks. The main strength of ACP+ is the avoidance of the choice of $\kappa$. %In the second part of the experiment, the tests of modified ACP+ have been made.

%\begin{figure}
%	\centering
	%\includegraphics[width=0.9\linewidth]{figures/AverageACP}
	%\vspace{-0.15in}
	%\caption{The empirical average $\overline{\Delta}_k$ results}
	%\label{average}
	%\vspace{-0.15in}
%\end{figure}

%\begin{figure}
%	\centering
%	\includegraphics[width=0.9\linewidth]{figures/MedianACP}
%	\vspace{-0.15in}
%	\caption{The empirical median $\overline{\Delta}_k$ results}
%	\label{median}
%		\vspace{-0.15in}
%\end{figure}

\begin{figure}
	\centering
    \includegraphics[width=0.84\linewidth]{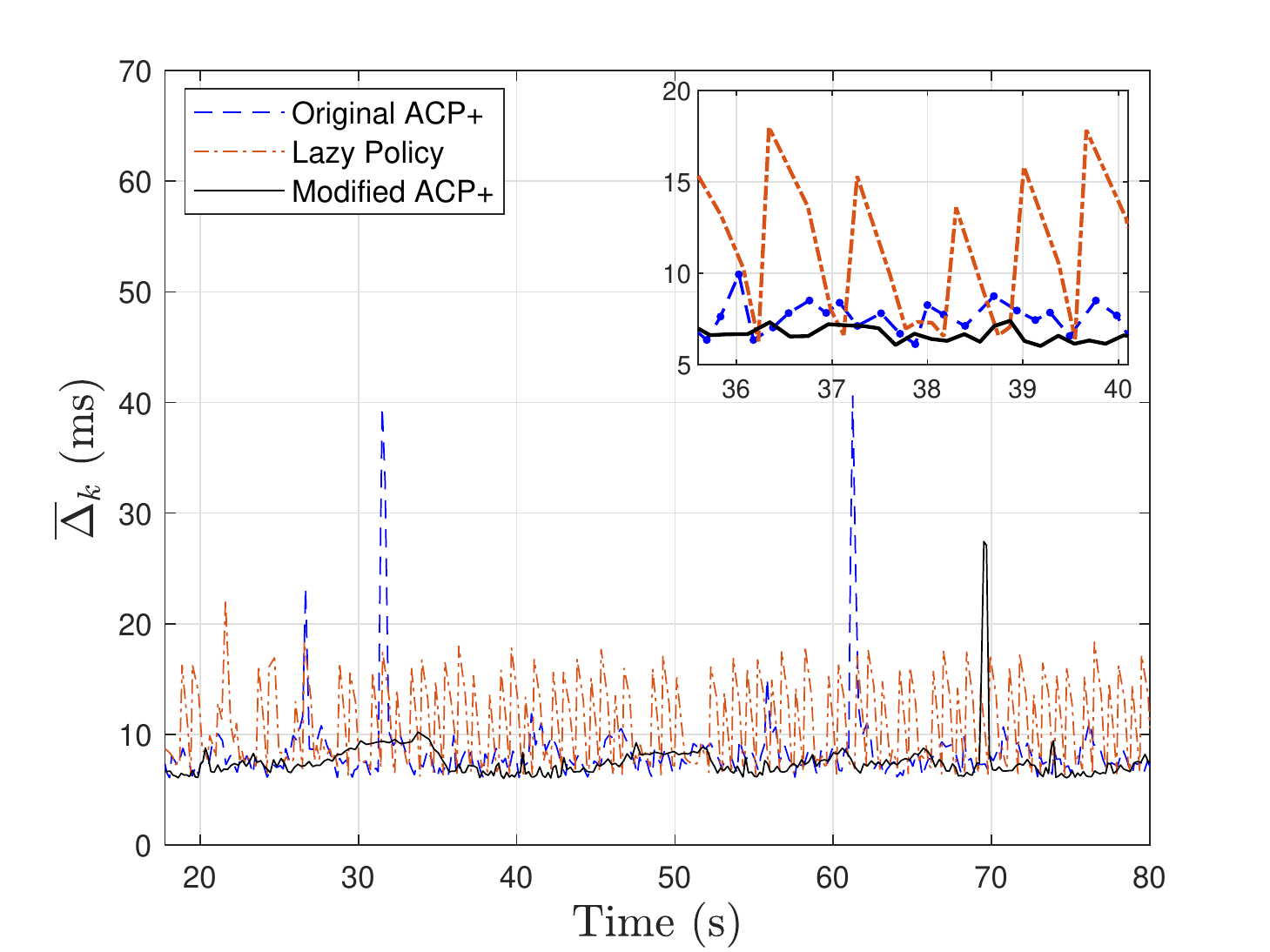}
%    \psfrag{Time (s)}{Plot of $\sin(t)$ and $\cos(t)$}
	\vspace{-0.1in}
	\caption{The trace of Lazy Policy and both versions of ACP+}
	\label{trace}
		\vspace{-0.2in}
\end{figure}

\begin{figure}
	\centering
	\includegraphics[width=0.84\linewidth]{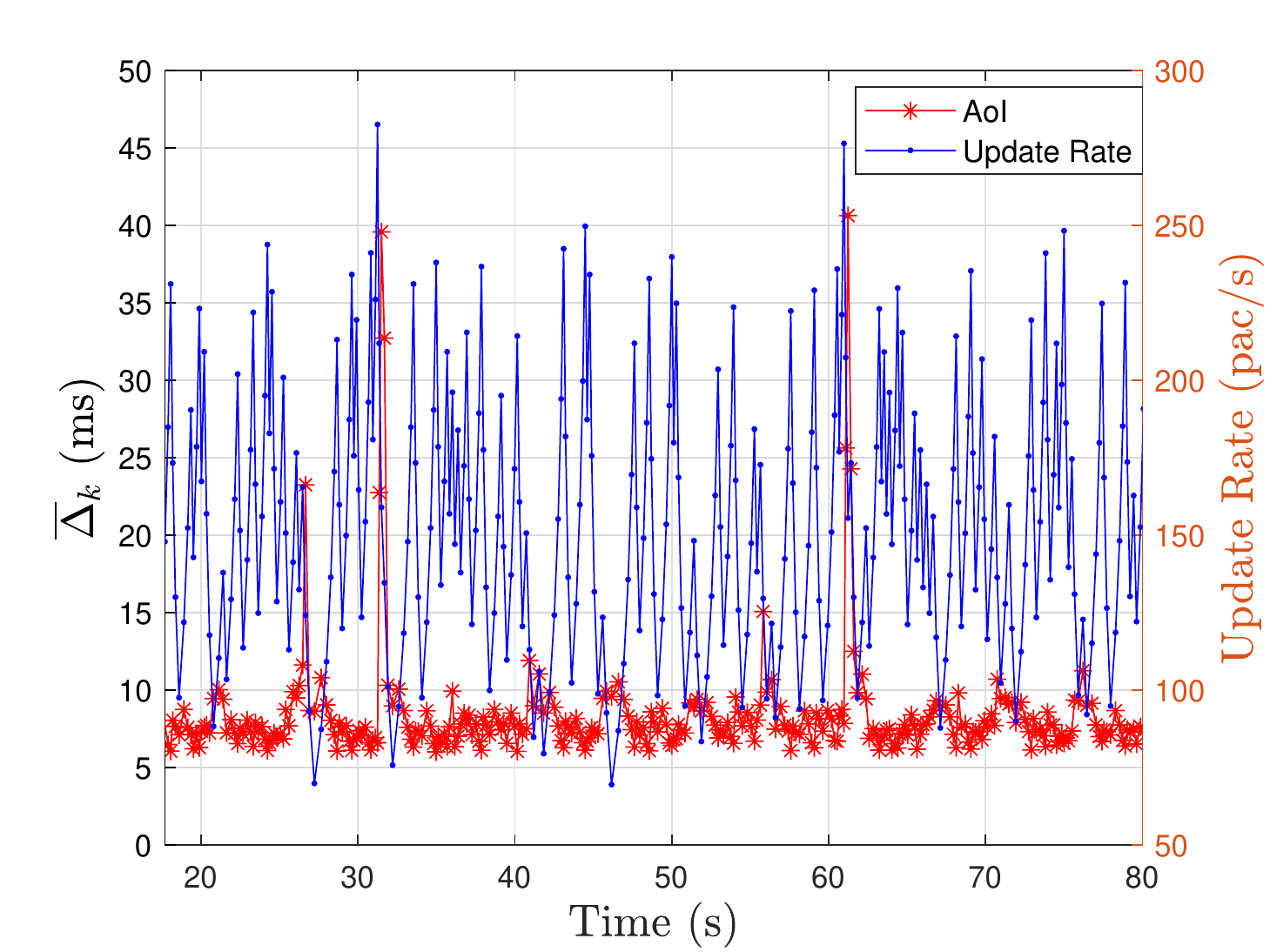}
	\vspace{-0.1in}
	\caption{AoI and update rate variations for original ACP+}
	\label{Peaks}
	\vspace{-0.15in}
\end{figure}

Fig. \ref{trace} shows the estimated $\overline{\Delta}_k$ values versus time graph for original ACP+, modified ACP+ and Lazy Policy. As seen in the figure, there are several peaks in both versions of ACP+. We notice that only the first peak of the original ACP+, blue line in Fig. \ref{trace}, occurs due to Issue 1. The second and third peaks exist since $\lambda_k$ is increased too much (see Fig. \ref{Peaks}), i.e., the clamping boundaries are too far. The reason for the peak in the modified one is over-increased $\lambda_k$ as well, but the age is relatively lower than the original ACP+. Modified ACP+ generally gives better results with low variance compared to the original ACP+ and Lazy Policy. The empirical variances of $\overline{\Delta}_k$ are calculated as 13.38, 20.93, and 3.24 for Lazy Policy, original ACP+, and modified ACP+, respectively. Top-right plot in Fig. \ref{trace} is a magnified part of the trace from $t=35.6$ sec to  $t=40.1$ sec.

\begin{figure}[ht]
\centering
\begin{subfigure}[b]{0.48\textwidth}
    \centering
	\includegraphics[width=0.84\linewidth]{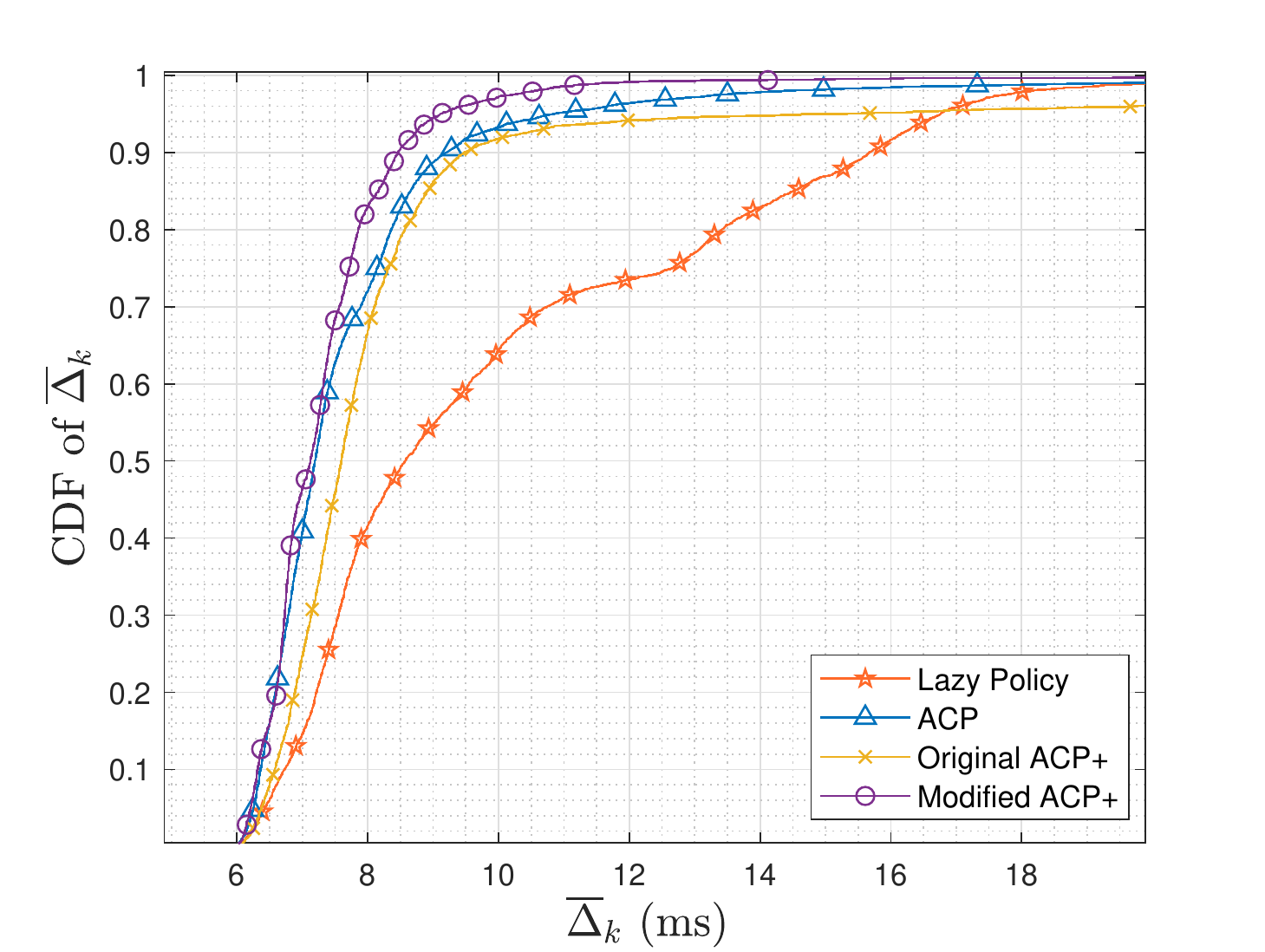}
	\vspace{-0.05in}
	\caption{The empirical AoI CDFs}
		\vspace{-0.2in}
	\label{acp+_cdf}
\end{subfigure}
	\vspace{-0.2in}
\begin{subfigure}[b]{0.48\textwidth}
    \centering
	\includegraphics[width=0.84\linewidth]{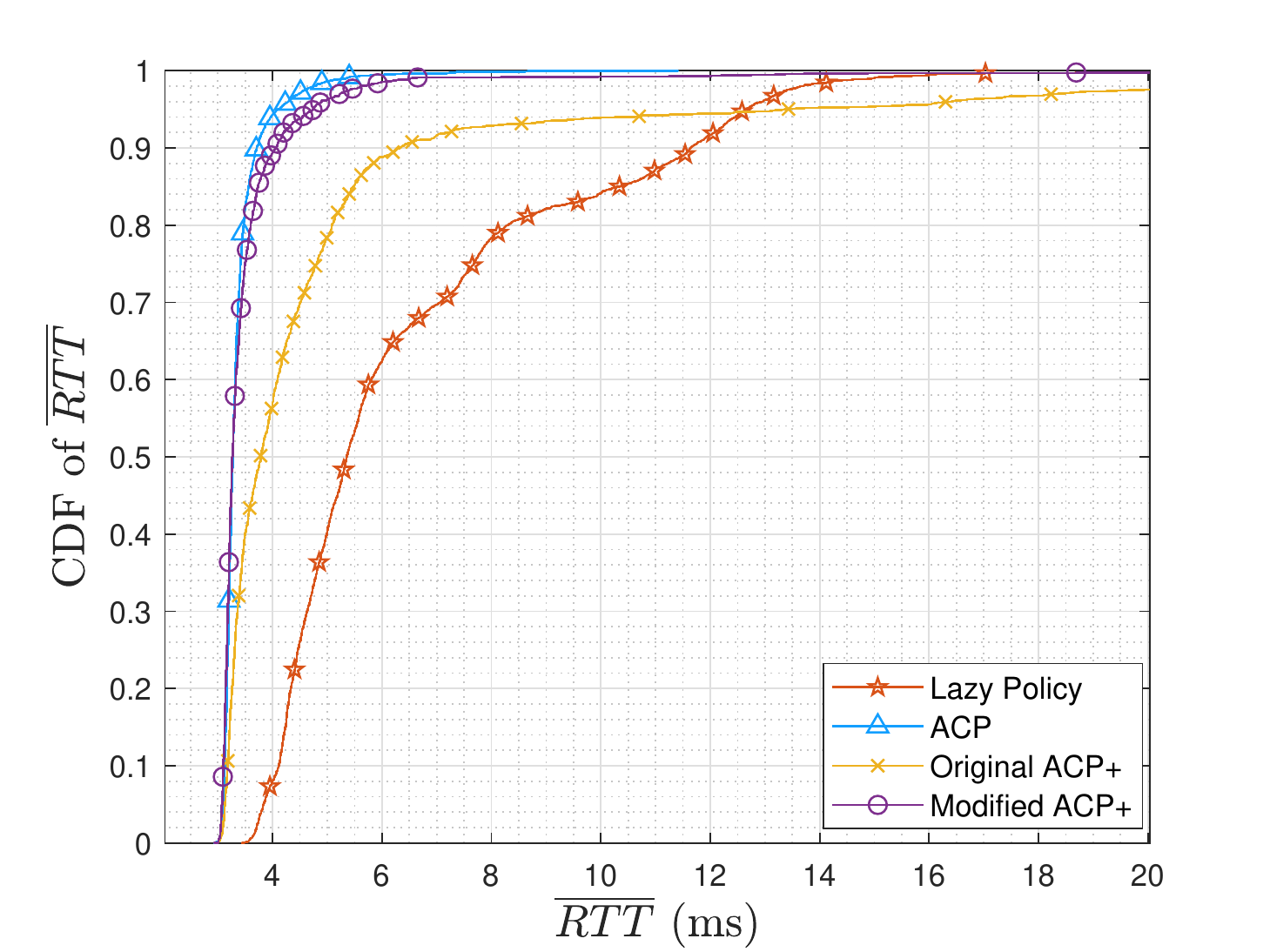}
	\vspace{-0.05in}
	\caption{The empirical RTT CDFs}
	\label{rtts}
\end{subfigure}
\caption{The empirical CDFs for Lazy Policy, ACP with $\kappa=0.1$ and two versions of ACP+}
	\vspace{-0.1in}
\end{figure}

Fig. \ref{acp+_cdf} and Fig. \ref{rtts} demonstrate the CDFs of $\overline{\Delta}_k$ and $\overline{RTT}$ for Lazy Policy, ACP with $\kappa=0.1$, and both versions of ACP+, respectively. There is an error margin in $\overline{RTT}$ because of Issue 1 explained in Sec. \ref{problem_solution}, the temperature of ESP32's, or other interfering signals. It is observed that the system did not encounter Issue 1 in ACP experiments with $\kappa=0.1$, but this problem is observed in ACP+ cases. Even though these problems may affect the results, it is expected that the differences in measured $\overline{RTT}$'s between the protocols mainly occur due to queueing delays in the hotspot and ESP32's. It can clearly be seen that modified ACP+ gives the best results in terms of AoI, whereas ACP gives the best results in terms of $\overline{RTT}$. One can deduce that modified ACP+ succeeded in minimizing AoI even if $\overline{RTT}$ is higher than ACP.

\begin{figure}
	\vspace{-0.1in}
	\centering
	\includegraphics[width=0.84\linewidth]{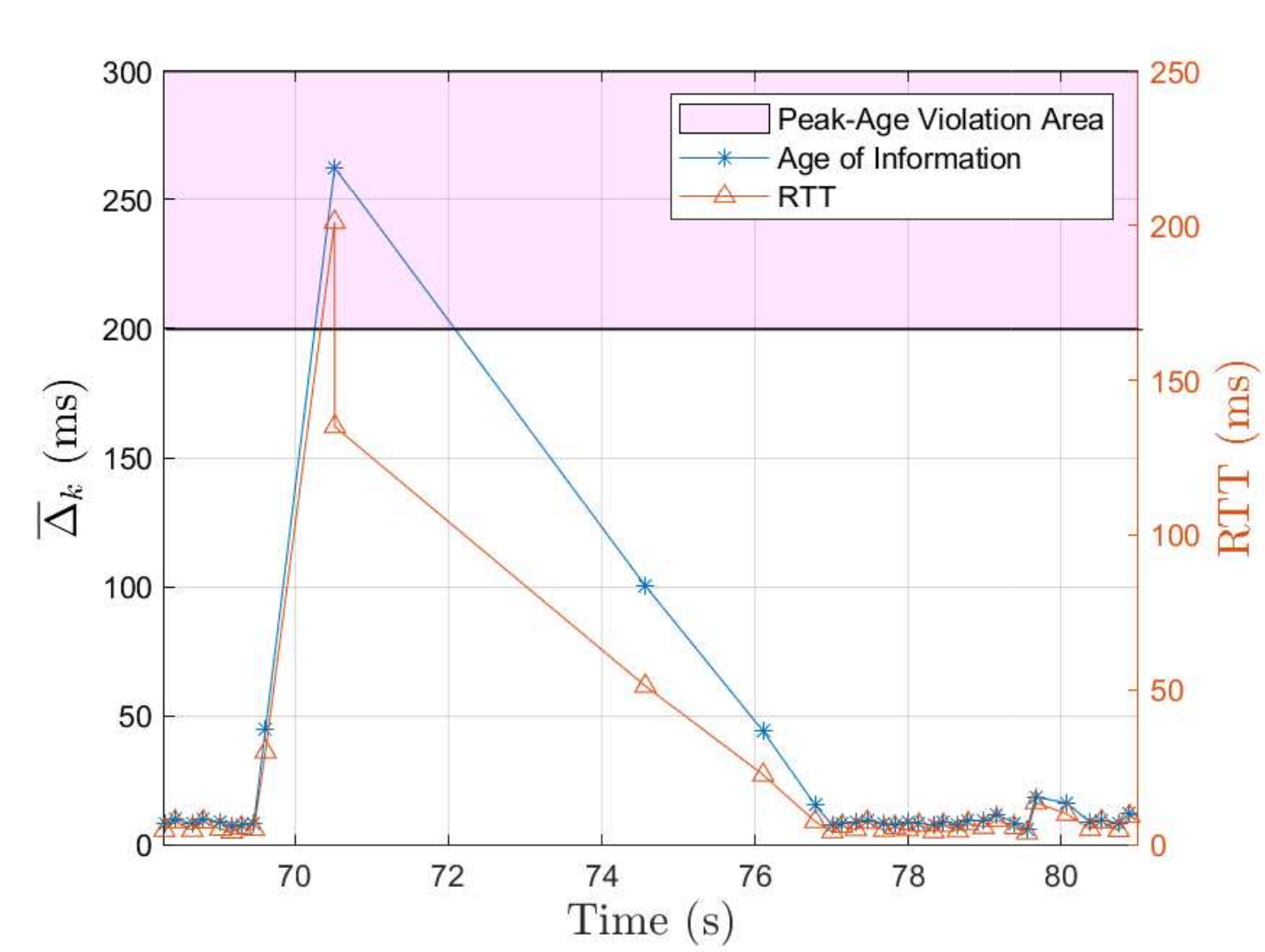}
	\vspace{-0.1in}
	\caption{Feedback Test}
	\label{Feedback_test}
		\vspace{-0.17in}
\end{figure}

We have conducted another test to measure the success of the designed feedback. Fig. \ref{Feedback_test} shows the trace of AoI and RTT values of ACP with $\kappa=0.5$. We set the peak age threshold as 200 ms.
When $t=t_v$, $\overline{RTT}$ is equal to 201.4 and $\overline{\Delta}_k$ is equal to 262.6. In other words, there is a peak age violation that triggers the feedback mechanism. After feedback activation, the decrease of $\overline{RTT}$ is observed. It decreases to 165.02 ms and triggers an increase on $\lambda_k$ and thereby decreasing $\overline{\Delta}_k$.

%In the figure, one can see the violation at $t=t_v$. At that time, $\overline{RTT}$ is equal to 201.4 ms before the mechanism is activated.

%Table \ref{tab:modify_tablo} presents a clear achievement of the modified version.
As stated in Solution 1, we multiplied $min(\overline{RTT},\overline{Z})$ by $30$ for ACP and $\frac{1}{\lambda_k}$ by $30$ for ACP+ to designate $\overline{T}$ so far, i.e., $\mathbf{\overline{T}_{30}}$ case is used. In $\mathbf{\overline{T}_{30}}$ case, we observe that modified ACP+ outperforms the original one in terms of average AoI (see Table 3). Lastly, we examine the effects of $\overline{T}$ over the age of information. One can see that the average age increases when (\ref{eq:epoch_length+}) is used to determine the length, which stands for the $\mathbf{\overline{T}_{10}}$ case for ACP+. The reason is that ACP+ updates $\lambda_k$ frequently and creates redundant oscillations in the $\mathbf{\overline{T}_{10}}$ case. Even though original ACP+ might have higher resiliency to network condition alterations, modified ACP+ offers prospering results for small-delay networks in both $\mathbf{\overline{T}_{10}}$ and $\mathbf{\overline{T}_{30}}$ cases.

%\begin{figure}%
%    \centering
%    \subfloat[\centering label 1]{{\includegraphics[width=3.950 cm]{figures/AverageACP.png} }}%
%    \qquad
%    \subfloat[\centering label 2]{{\includegraphics[width=3.950 cm]{figures/MedianACP.png} }}%
%    \caption{2 Figures side by side}%
%    \label{fig:example}%
%\end{figure}

\begin{table}
\vspace{-0.15in}
\centering
\caption {The estimated average AoI values}
\vspace{-0.1in}
\normalsize
\begin{tabular}{l|c|c|}
\cline{2-3}
 & \multicolumn{1}{l|}{Original ACP+} & \multicolumn{1}{l|}{Modified ACP+} \\ \hline
\multicolumn{1}{|l|}{$\mathbf{\overline{T}_{10}}$} & 9.15 ms & 8.93 ms \\ \hline
\multicolumn{1}{|l|}{$\mathbf{\overline{T}_{30}}$} & 9.01 ms & 7.54 ms \\ \hline
\end{tabular}
\label{tab:modify_tablo}
\end{table}

\vspace*{-0.1cm}
\vspace{-0.01in}
\section{Conclusions}
\label{conclusion}

This study evaluated ACP and ACP+ \cite{tanya} on IoT devices, specifically ESP32's. We have reported several issues related to applying ACP and ACP+ in this setting and suggested solutions to solve these issues. A feedback mechanism was designed to exit from the loop which ACP and ACP+ enter due to ESP32's buffer management error.  AoI measurements in real-world networks under ACP for different $\kappa$ values, the Lazy Policy in \cite{tanya} and ACP+ have been conducted. Results indicate that the original ACP+ is too greedy for small-delay networks. A modified version of ACP+ was designed to solve this problem of the original ACP+, and results show that modified ACP+ gives superior results. As future work, we plan to make clamping boundaries RTT-dependent such that ACP+ become compatible and deployable for every IoT device in every network. 
\vspace{-0.0101in}

% Not modifiable, (variable) 

\ifCLASSOPTIONcaptionsoff
  \newpage
\fi

\vspace*{-0.75cm}

\section{Acknowledgements}

This study has been supported by TUBITAK grants 117E215 and 119C028. We also would like to thank Tanya Shreedhar for discussions.

% Add Huawei.

\vspace*{-0.05cm}
\bibliographystyle{IEEEtran} 
\bibliography{combined}

\end{document}